\documentclass[twocolumn]{aastex631}

\usepackage[english]{babel}
\usepackage[utf8]{inputenc}

\usepackage{amssymb}
\usepackage{hyperref}
\usepackage{lipsum}
\usepackage{amsmath}
\usepackage{graphicx}

\usepackage{natbib}

\newcommand{\REV}[1]{\textcolor{black}{#1}}

\shorttitle{A forecasting framework for galactic cosmic ray flux in space weather applications}
\shortauthors{Pelosi et al.}

\begin{document}

\title{A forecasting framework for galactic cosmic ray flux in space weather applications}
\submitjournal{Advances in Space Research}


\correspondingauthor{David Pelosi}
\email{david.pelosi@pg.infn.it}

\author[0009-0003-4663-1262]{David Pelosi} 
\affiliation{Università degli Studi di Perugia}
\affiliation{INFN - Sezione di Perugia}

\author[0000-0002-8346-9941]{Fernando Barão} 
\affiliation{Laboratório de Instrumentação e Física Experimental de Partículas}

\author[0000-0001-7584-293X]{Bruna Bertucci} 
\affiliation{Università degli Studi di Perugia}
\affiliation{INFN - Sezione di Perugia}

\author[0000-0002-9448-846X]{Francesco Faldi} 
\affiliation{Università degli Studi di Perugia}
\affiliation{INFN - Sezione di Perugia}

\author{Emanuele Fiandrini} 
\affiliation{Università degli Studi di Perugia}
\affiliation{INFN - Sezione di Perugia}

\author[0000-0002-9127-7608]{Alejandro Reina Conde} 
\affiliation{INFN - Sezione di Bologna}
\affiliation{Centro Nacional de Aceleradores (CNA)}

\author[0000-0003-1874-2144]{Miguel Orcinha} 
\affiliation{INFN - Sezione di Perugia}

\author[0000-0002-0856-9299]{Nicola Tomassetti} 
\affiliation{Università degli Studi di Perugia}
\affiliation{INFN - Sezione di Perugia}

\begin{abstract}
The intensity and energy spectrum of galactic cosmic rays in the heliosphere are significantly influenced by the 11-year solar cycle,a phenomenon known as solar modulation. Understanding this effect and its underlying physical mechanisms is essential for assessing radiation exposure and associated risks during space missions.
Starting from a previously developed effective predictive model of solar modulation, validated using cosmic-ray flux measurements from space-based detectors such as PAMELA and AMS-02, we build a generalizable forecasting strategy for the long-term evolution of cosmic-ray fluxes. This strategy is based on identifying delayed cross-correlation relationships between solar proxies and the model’s parameters. It integrates recent findings on time lags between cosmic ray fluxes and solar activity, and incorporates advanced time-series signal processing techniques.
The framework not only performs well in reproducing observed data, but also shows strong potential for applications in space radiation monitoring and forecasting. By efficiently capturing the long-term variability of galactic cosmic rays, our approach contributes valuable insights for evaluating radiation risks, ultimately supporting safer and more effective space exploration.
\end{abstract}


\section{Introduction}
Galactic cosmic rays (GCRs) traversing the heliosphere undergo intricate interactions with solar magnetic fields and disturbances in the solar wind, resulting in fluctuations in their intensity and energy spectra. This phenomenon, referred to as solar modulation, significantly alters the GCR flux in the inner heliosphere in comparison to the Local Interstellar Spectrum (LIS), which is the flux outside the heliopause that is assumed to remain constant over time. Solar activity follows an 11-year cycle, which is reflected in changes in solar proxies such as the sunspot number (SSN). This cycle strongly correlates with the modulation effect on cosmic rays. 
\REV{During the maximum of solar activity, increased solar magnetic field strength and turbulence can be measured near Earth. These correlate to greater modulation of galactic particles, causing the flux near Earth to diminish.} Conversely, during solar minimum, the Sun's shielding effect becomes less pronounced. This cyclic variability highlights the intricate interplay between solar activity and the modulation of cosmic rays within the heliosphere. Understanding this phenomenon is crucial for unraveling the origins of cosmic particles and antiparticles, as well as for identifying anomalies that could unveil new astrophysical sources. In this respect, numerous models have been developed with the aim of predicting the evolution of GCR intensity \citep{Norbury2018}.
Moreover, comprehending the dynamics of the transport of cosmic rays in the heliopshere is essential for evaluating radiation doses and risks to spacecraft electronic components and crew health during space missions. This underscores the significance of accurately predicting cosmic ray fluxes both near Earth and in interplanetary space \citep{cucinotta, Papadopoulos, Alia}.
To comprehend the dynamics of cosmic ray modulation and its correlation with solar activity, studying the time lag between the two phenomena is essential. Several studies have reported a lag of a few months between the monthly SSN and corresponding variations in neutron monitor rates, indirectly measuring GCR fluxes on Earth \citep{Lag_General}.
\REV{While there is no clear consensus on the origin of the time lag, it is often interpreted as relating to the time required for heliospheric plasma to propagate outward and influence the modulation region, and to the typical trajectories charged particles take while propagating through the solar system. However, the observed lag likely results from a complex interplay of several fundamental processes governing the transport of charged particles: gradient and curvature drifts of GCRs in the large-scale interplanetary magnetic field, convection with the latitude-dependent solar wind, and rigidity-dependent diffusion through magnetic irregularities in heliospheric turbulence. Additionally, the lag may be influenced by delayed responses of GCRs to evolving heliospheric conditions, or by delays in the formation of the heliospheric magnetic field in response to sunspot activity \citep{Allen_Lag,Dorman_Lag, Wang_Lag}.}
\subsection{\REV{Experimental datasets}}
Numerous experiments have provided invaluable data essential for studying cosmic ray modulation dynamics. Ground-based observatories and space probes continuously monitor solar activity, offering crucial insights into the temporal evolution of proxies that control the solar cycle evolution. Notable contributions have been made by Voyager 1 and Voyager 2 probes, which have been providing crucial data since crossing the heliopause in mid-2012 and late 2018, respectively, aiding in constraining LIS models \citep{Cummings_2016,Stone2019}. 
Additionally, GCR flux measurements have been conducted using various instruments over the years. Notable examples include the Electron Proton Helium Instrument (EPHIN) onboard the Solar and Heliospheric Observatory (SOHO) \citep{SOHO}, detectors mounted on balloons in the upper atmosphere such as BESS \citep{BESS_04_07}, \citep{BESS97}, and space-borne instruments like PAMELA \citep{PAMELA_first}. More recently, continuous measurements from spectrometers like the Cosmic Ray Isotope Spectrometer (CRIS) on the ACE spacecraft \citep{ACE/CRIS_first}, and advanced instruments like AMS-02 \citep{AMS02_first} have provided crucial data. These  multi-channel datasets, along with a precise assessment of the lag, offer unprecedented opportunities to refine our understanding of cosmic ray transport dynamics enhancing our ability to develop robust predictive models for GCR fluxes in the heliosphere.
\subsection{\REV{Modeling approach}}
In this context, we used an effective and predictive model of solar modulation. Our model incorporates fundamental physics processes of particle transport, such as diffusion, advection, and adiabatic cooling, to compute the energy spectrum and temporal evolution of cosmic radiation in the inner heliosphere. We utilize a numerical approach to solve the Parker equation in its radial approximation and develop a forecasting strategy that relates the model parameters to solar activity proxies, accounting for the time lag between these two quantities. This allows us to estimate cosmic-ray fluxes in advance, based exclusively on solar activity conditions.
\subsection{\REV{Outline}}
This paper is organized as follows: in Section \ref{sec:LIS}, we introduce the model used to derive the LIS by integrating data from the Voyager probes and AMS-02. Section \ref{sec:calculations} discusses the physics of GCRs propagation in the heliosphere, introducing the radial Parker equation and the numerical method used to solve it.
Section \ref{sec:calib} presents the calibration procedure based on proton data, the fitting procedure, and demonstrates the model's performance in reconstructing spectra for different epochs.
Sections \ref{sec:temporal_evolution}, \ref{sec:EMD} and \ref{sec:cross_correlation} detail the process of obtaining time-series of the free parameter of the model, as well as the procedure for smoothing and extracting correlations between model parameter and solar proxy.
Section \ref{sec:Results} showcases some forecasting results, comparing our models with validation datasets from various experiments for different GCR species and energy ranges.
Section \ref{sec:conclusion} concludes by summarizing and discussing the model's capabilities, limitations, areas for improving predictions, and potential future applications.

\section{Methods}

\subsection{\REV{Determination of the Local Interstellar Spectra}}
\label{sec:LIS}
\begin{figure}[!ht]
    \centering
    \includegraphics[width=0.5\textwidth]{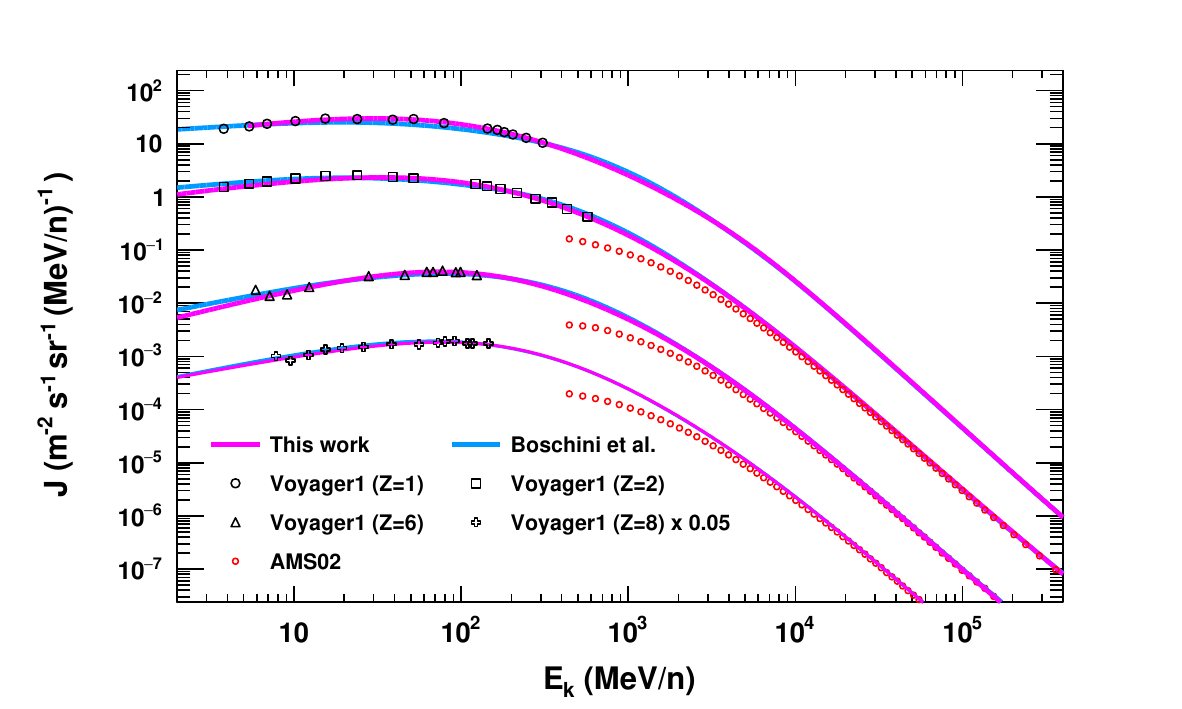}
    \caption{Comparison among LIS models representing different GCR species such as protons, He, C, and O, with the magenta line indicating the LIS utilized in this study \citep{reinaconde2022}. Low-energy data from Voyager \citep{Cummings_2016} and high-energy data from AMS-02 \citep{AMSPReports2021}, \citep{AMSPRL2017} are also included. The Oxygen LIS and data are scaled by an arbitrary factor for visualization purposes.}
    \label{fig:LIS}
\end{figure}

The knowledge of the interstellar flux of cosmic rays is crucial for studying solar modulation and effects such as variations in the flux ratios among different species. Understanding conventional astrophysical backgrounds and the local galactic environment is essential for having a comprehensive understanding of phenomena ranging from radio and gamma-ray emissions to the physics of the interstellar medium. This knowledge is key to extracting possible weak signals of new physics, such as dark matter annihilation or decay.
Until the end of 2012, it was not possible to measure the LIS in situ. With the Voyager probes crossing the outer heliosphere, data became available, and measurements of the LIS below $\sim500\,$MeV were obtained. Moreover, space-borne instruments measure the GCR flux near Earth, covering the largely unmodulated high-energy part of the spectrum above $10$ GeV. Specifically AMS-02 proton and helium fluxes provide valuable information, allowing us to elucidate the shape of the LIS.
These data are crucial and can be used to develop parametric models \citep{LIS_Herbst, LIS_NM} or spatially dependent models \citep{Tomassetti_Donato, feng}, derived from a full calculation approach by solving cosmic ray propagation equations within the Galaxy. Additionally, recent LIS models have been derived by integrating multi-channel data with state-of-the-art galactic propagation codes \citep{GALPROP, LIS_Boschini_2020}.
In this work, we used the LIS model previously developed in \citet{reinaconde2022}, which extends the method described in \citet{LIS_Corti_2016}. The LIS is parameterized as a combination of four smooth power laws in rigidity, expressed as:

\begin{equation}
\REV{
J_{\mathrm{LIS}}(P)=N\left({\frac{P}{1\,\text{GV}}}\right)^{\gamma_{0}}\prod_{i=1}^{3}\left[{\frac{1+(P/P_{i})^{s_{i}}}{1+P_{i}^{-s_{i}}}}\right]^{\Delta_{i}/s_{i}}
}
\end{equation}

where $N$ is the flux normalization at $1\,$GV, $\gamma_0$ is the spectral index of the first power law, $\Delta_i = \gamma_i - \gamma_{i-1}$ denotes the difference in spectral index between the i-th power law and the previous one, $P_i$ indicates the rigidities at which the breaks between power laws occur and $s_i$ controls the smoothness of the breaks.
The data-driven approach relies on specific datasets necessary to fit the model. Voyager-1 conducted exclusive direct measurements of the proton LIS spectrum after crossing the heliopause, particularly within the energy range of 140-320 MeV. Meanwhile, for the high-energy component of GCRs unaffected by modulation, energy-resolved data from AMS-02 \citep{AMSPReports2021} serve as appropriate inputs for defining the LIS. To account for residual effects of solar modulation, we utilize Force-Field approximation \citep{FFA_ref} \REV{based on the determination of modulation potential $\phi$}. Altogether, we require twelve parameters to determine our data-driven LIS model: eleven from the LIS parameterization and one from the Force-Field approximation. We calculate these twelve parameters through a $\chi^2$ minimization using MINUIT2 \citep{MINUIT2} routines.
Thanks to the extremely high precision nuclei spectra recently measured by AMS-02 \REV{\citep{AMS_nuclei}}, integrated with Voyager-1 data, the procedure has also been repeated to obtain the LIS for Helium, Carbon, and Oxygen.
Figure \ref{fig:LIS} shows a comparison between different models of the LIS discussed above for different species of GCRs. As shown, the high degree of agreement is primarily due to the recent available datasets constraining the models.
\REV{The corresponding parameter values for each species are listed in Table \ref{tab1}}.

\begin{table*}[t]
\begin{tabular}{l
                @{\hskip 10pt} r@{\,$\pm$\,}l
                @{\hskip 10pt} r@{\,$\pm$\,}l
                @{\hskip 10pt} r@{\,$\pm$\,}l
                @{\hskip 10pt} r@{\,$\pm$\,}l}
\hline
 & \multicolumn{2}{c}{p} & \multicolumn{2}{c}{He} & \multicolumn{2}{c}{C} & \multicolumn{2}{c}{O} \\
\hline
$\phi$ (MV) & 578   & 9     & 574   & 8     & 590   & 9     & 606   & 8     \\
$N$ ($\mathrm{m^{-2}\ sr^{-1}\ s^{-1}\ GV^{-1}}$)
           & 5396  & 72    & 407.8 & 5.0   & 8.418 & 0.079 & 8.227 & 0.087 \\
$\gamma_0$ & 1.889 & 0.005 & 1.920 & 0.005 & 2.596 & 0.004 & 2.734 & 0.005 \\
$P_{0}$ (MV) & 0.494 & 0.005 & 1.026 & 0.007 & 0.982 & 0.006 & 1.189 & 0.008 \\
$s_0$ & 1.712 & 0.021 & 1.792 & 0.018 & 2.178 & 0.019 & 2.387 & 0.031 \\
$\Delta_0$ & -4.253 & 0.005 & -4.064 & 0.005 & -4.365 & 0.004 & -4.128 & 0.005 \\
$P_{1}$ (MV) & 5.847 & 0.017 & 2.205 & 0.009 & 3.321 & 0.059 & 6.711 & 0.221 \\
$s_1$ & 3.471 & 0.034 & 0.927 & 0.043 & 1.278 & 0.044 & 0.277 & 0.009 \\
$\Delta_1$ & -5.645 & 0.073 & -0.705 & 0.006 & -1.022 & 0.005 & -2.222 & 0.009 \\
$P_{2}$ (MV) & 759.9 & 4.4   & 736.4 & 1.6   & 601   & 258   & 709   & 71    \\
$s_2$ & 1.527 & 0.033 & 0.873 & 0.074 & 1.473 & 0.239 & 0.859 & 0.035 \\
$\Delta_2$ & 0.414 & 0.161 & 0.457 & 0.056 & 0.349 & 0.083 & 1.069 & 0.058 \\
\hline
\end{tabular}
\vspace{0.5em}
\caption{\REV{LIS parameters values and relative errors for proton, helium, carbon and oxygen, following the minimization process described in Section \ref{sec:LIS}. Here $\phi$ is the modulation parameter/potential of the FFA, $N$ is the flux normalization at 1~GV, $\gamma_0$ is the spectral index of the first power law, $\Delta_i = \gamma_i - \gamma_{i-1}$ is the difference in spectral index between the $i$-th power law and the previous one, $P_i$ are the rigidities at which the breaks between power laws happen, and $s_i$ controls the smoothness of the breaks.}}
\label{tab1}
\end{table*}

\subsection{Numerical Solution of the Parker Equation}
\label{sec:calculations}
The transport of GCRs in the heliosphere is described by the Parker equation \citep{PARKER19659}. This equation governs the evolution of the phase-space density $\psi = \psi(\vec{r},P,t)$ for a given particle specie:
\begin{align}
\label{eq:parker}
\frac{\partial\psi}{\partial t} =\; & \nabla\cdot\left(\textbf{K}_s\cdot\nabla\psi\right) 
- (\textbf{V}_{sw} + \langle\textbf{V}_{D}\rangle) \cdot\nabla\psi \nonumber \\
& + \frac{1}{3} (\nabla \cdot  \textbf{V}_{sw}) \frac{\partial\psi}{\partial\ln P} + Q
\end{align}
where $P$ represents rigidity. The terms in the equation correspond to advection with solar wind speed $\textbf{V}_{sw}$ and pitch-angle-averaged drift velocity $\langle\textbf{V}_{D}\rangle$, diffusion defined by the symmetric part of the diffusion tensor $\textbf{K}_s$, adiabatic momentum losses, and $Q$ as GCR sources.
Analytic solutions of this equation rely on simplifying approximations, particularly concerning diffusion or losses, such as Force-Field and Convective-Diffusion approximation \citep{NTPRL}. The Force-Field approach describes the impact of solar modulation on observed cosmic rays by reducing the kinetic energy of charged particles through a modulation potential $\phi$ typically ranging from $0.1$ to $1$ GV. Conversely, the Convective-Diffusion calculations offers a simplified solution to the Parker equation, neglecting spatial dependencies and considering only the balance between diffusive motion and adiabatic cooling.
Since the focus of this work is to develop a forecasting strategy based on polarity-dependent cross-correlation between free modulation parameters and a solar proxy, we decided to use a simple and well-known solution of the Parker equation, based on a numerical model employing the Crank–Nicolson implicit scheme. Assuming spherical symmetry, a radially flowing solar wind with velocity $V$, and an isotropic diffusion tensor, the Parker transport equation simplifies under steady-state conditions ($\partial\psi / \partial t = 0$) and without source terms ($Q = 0$), resulting in a second-order parabolic equation in space and first-order in rigidity:
\begin{align}
\label{eq:Parker_radial}
K\frac{\partial^{2}\psi}{\partial r^{2}} 
+ \left( \frac{\partial K}{\partial r} + \frac{2K}{r} - V \right)\frac{\partial\psi}{\partial r} 
& \nonumber \\
+ \left( \frac{2V}{3r} + \frac{1}{3} \frac{\partial V}{\partial r} \right)\frac{\partial\psi}{\partial\ln P} 
&= 0
\end{align}
\REV{where $\psi$ relates to the isotropic cosmic ray flux ($J$) in kinetic energy per nucleon ($T$) as $J(T) = P^2\,\psi(P)$ \citep{Moraal2013}}.
The computational domain is discretized into a grid of $\ln P \times r$, with boundary conditions set at $r = r_{\odot}$ (about $1\%$ of the distance between the Sun and Earth) and $r_{HP} = 122\,$AU (the heliopause, where the flux matches the LIS) while the rigidity range is set between $P_{min} = 50\,$MV and $P_{max} = 1\,$TV. The grid consists of $N = 610$ radial nodes and $M = 500$ rigidity nodes.
The partial differential equation is transformed into a system of linear equations, solved iteratively as detailed in \citet{NTPRL}.
This approach allows for testing various forms of the effective radial diffusion coefficient $K$ which incorporates both the parallel and perpendicular diffusion coefficients, including mass/charge-dependent effects through its dependence on rigidity and velocity. The benchmark diffusion coefficient adopted in this work is given by:
\begin{equation}
    K(P,t) = K_0(t)\, \beta(P)\, \frac{P}{P_0}
    \label{eq:diffusion_coeff}
\end{equation}
The mass/charge dependence of $K$ is embedded in the $\beta$ factor, $K_0$ is the normalization factor, typically of the order $\sim 10^{22}\mathrm{cm^2 \, s^{-1}}$, referenced at $P_0 = 1$ GV. 
For simplicity, no radial dependence is included in $K$ in this model. Temporal variations in $K$ reflect changes in global heliospheric conditions, represented by the evolving modulation parameter $K_0(t)$.
\REV{It is worth noting that the model exhibits a strong degeneracy between the solar wind speed $V$, the diffusion normalization $K_0$, and the size of the modulation boundary, such that only their ratios affect the solution. Since the model is calibrated using data near Earth, any change in $V$ or boundary location can be exactly compensated by a rescaling of $K_0$, without affecting the solution at Earth. To eliminate this redundancy and ensure numerical stability, we fix $V = 450\,\mathrm{km/s}$ and place the boundary at the heliopause, leaving $K_0$ to be fitted. }

\subsection{Model Calibration}
\label{sec:calib}
The calibration of the model parameter relies on time- and energy-resolved proton data from PAMELA and AMS-02. Specifically, we use the daily fluxes measured by the AMS-02 experiment from May 2011 to November 2019 \citep{AMS_PRL2021}, Bartel's rotation-averaged (BR)\footnote{Bartel's rotation = 27 days}, and the Carrington rotation-averaged fluxes\footnote{Carrington rotation = 27.27 days} observed by the PAMELA instrument operating on the Resurs-DK1 satellite from June 2006 to January 2014 \citep{PAMELA_2013_protons, PAMELA_2018_protons}.
As discussed in Section \ref{sec:calculations}, we adopt a quasi-steady approach, wherein a time-series of steady-state solutions for GCR flux corresponds to a time-series of input parameter $K_0$. To ensure the validity of this approximation, we employ monthly resolution for the calibration data. This holds because the characteristic timescale of the periodicities associated with solar modulation does not exceed the monthly resolution at the rigidity typical of AMS-02 and PAMELA data. Typically, these timescales cover a wide range of hours to years and generally decrease with increasing rigidity, as described by \citet{Strauss}, despite some unusual features recently observed by the AMS-02 collaboration \citep{AMS02_Temporal_Structures}.
The advantage of using calibration data spanning from 2006 to 2019 is that it covers different solar activity conditions over two different solar cycles (23rd and 24th). This period includes the solar minimum occurring during an epoch of the negative polarity phase ($A < 0$) from 2006 to 2009, the ascending phase to the solar maximum around February 2014, and the following descending phase in $A > 0$ phase.
It is worth noting that during any maximum the polarity of the heliospheric magnetic field (HMF) reverses, resulting in an undefined polarity condition lasting several months due to strong turbulence in the solar magnetic field, which no longer identifies as a dipole. 
How to deal with the reversal will be presented later on.
The PAMELA and AMS-02 data cover an extended energy range from $88\,$MeV to $46.5\,$GeV and from $0.45\,$GeV to $100\,$GeV, respectively. This extensive range is essential for accurately determining the best-fit parameters of the model via fitting procedures.

\begin{figure*}[!t]
    \centering
    \includegraphics[width=0.8\textwidth]{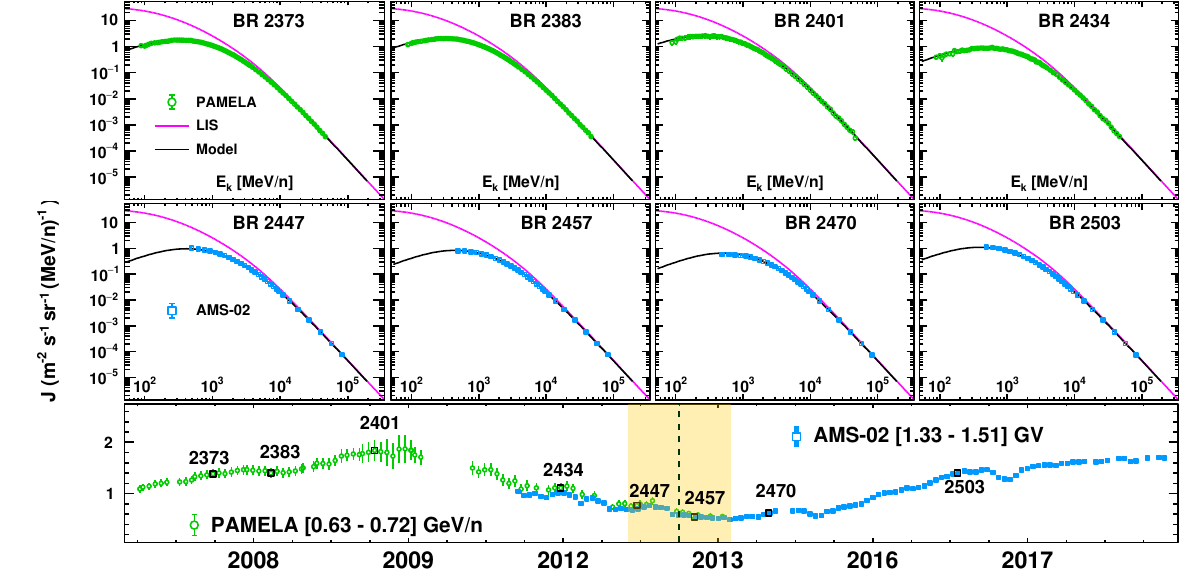}
    \caption{Modulated energy spectra measured by PAMELA and AMS-02 over 8 selected Bartel rotations (BR), marked as black boxes in the bottom panel. The calculations show the best-fit proton fluxes as black lines, while the magenta line represents the LIS used in this work. The model demonstrates good consistency with the data across all explored epochs. The bottom panel shows the temporal evolution of the proton flux measured by PAMELA and AMS-02 for two reference energy bins. The yellow shaded band highlights the reversal phase, as detailed in \citet{Sun_reversal_2013}.}
    \label{fig:model_performance}
\end{figure*}

It's worth noting that $K_0$ is degenerate with the constant solar wind speed $V$. Model's sensitivity is to the combination $K_0/V$, which allows for compensation through proper rescaling of $K_0$. Thus, the wind speed $V$ remains fixed at a reference value of $450\,$km/s.
The performance of the model for 8 different BRs is illustrated in Fig.\ref{fig:model_performance}, where the black lines represent our calculations for the best-fit proton fluxes, while the LIS used in the calculation is shown by the magenta line. Fits were independently conducted for each BR using the standard MINUIT routine. The minimizing function for a given BR is defined as:
\begin{equation}
\chi^{2}(K_0) = \sum_{j} \,\frac{1}{\sigma_{j}^2} \, \left[ \, \hat{J}_\mathrm{p,\, j} - \frac{1}{\Delta P_j} \int_{P_{j}}^{P_{j+1}} J_{\mathrm{p}}(P,K_0) \, dP \, \right]^{2}
\end{equation}
\vspace{0.1em}

where an integration of the model output $J_\mathrm{p}(P,K_0)$ is computed within each rigidity bin $\Delta P_j$, while $\hat{J}_\mathrm{p,\, j}$ is the mean value of the data. Factors $\sigma_j$ account for the experimental uncertainties in the measured fluxes.
The results demonstrate that, despite the approximations, the model effectively reconstructs the cosmic ray flux during any phase of solar activity.
Fits for each BR (or Carrington Rotation) yield a time-series of best-fit values for the parameter $K_0$, reflecting the temporal evolution of the GCR modulation as described in Section \ref{sec:temporal_evolution}.

The  best-fit results of the parameter $K_0$ from the time-resolved proton flux measurements of PAMELA and AMS-02 are reported with green and blue markers respectively, in blue the same parameters are reported after smoothing with the EMD approach described in the text (refer to Sect. \ref{sec:EMD}). As a reference of the solar activity conditions corresponding to the used dataset, the temporal evolution of smoothed SSN data is also presented in the bottom-most plot. The vertical dashed line represents the reversal epoch $T_{rev}$, while the shaded area around it indicates the transition epoch where the HMF polarity is not defined.

\subsection{Temporal Evolution}
\label{sec:temporal_evolution}
\begin{figure}
\vspace{0.4cm}
    \centering
    \includegraphics[width=0.5\textwidth]{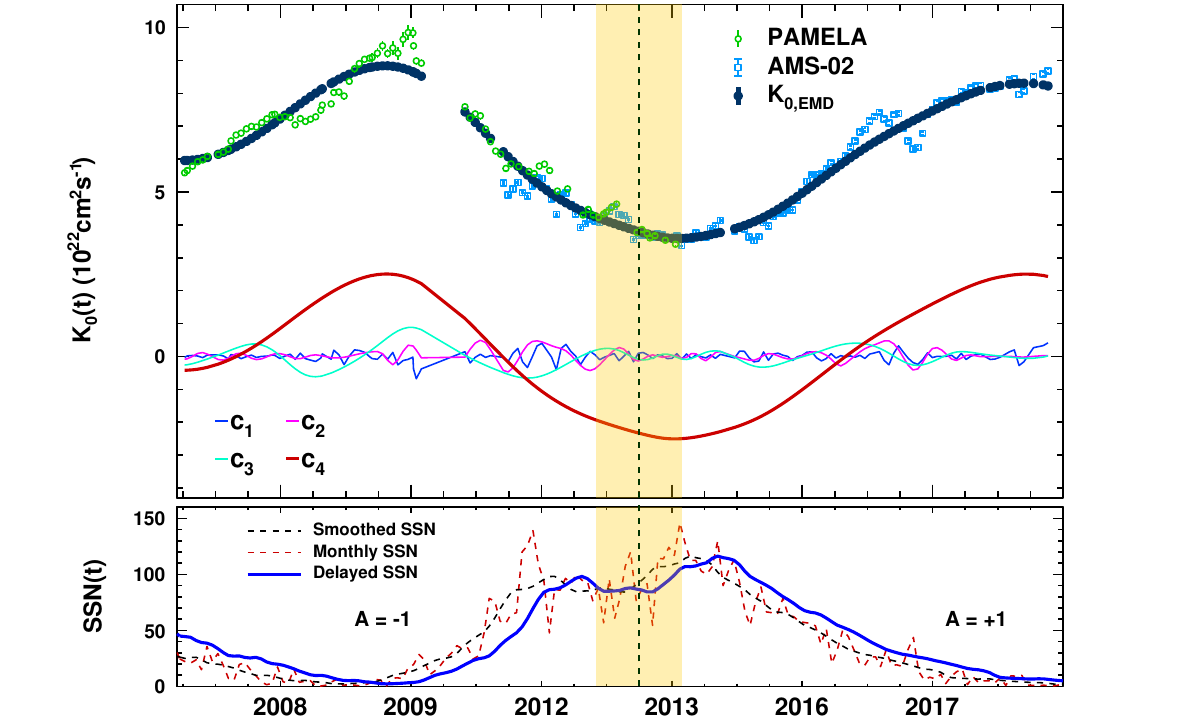}
\caption{The best-fit results of the model parameter $K_0$ from the time-resolved proton flux measurements of PAMELA and AMS-02 are reported in green and blue markers respectively. \REV{The dark blue points represent the same parameter after smoothing using the EMD approach described in the text (see Sect. \ref{sec:EMD}), whose intrinsic mode functions $c_i$, with $i = 1, \dots, 4$, are also displayed. Relating to the heliospheric conditions during the period under study, the bottom panel shows the temporal evolution of the SSN in three regimes: monthly values (dashed red), 13-month moving average smoothed values (dashed black), and the delayed smoothed SSN (solid blue) according to the time lag defined in Equation \ref{eq:lag}}. The vertical dashed line represents the reversal epoch $T_{rev}$, while the shaded area around it indicates the transition epoch where the HMF polarity is not defined.}

    \label{fig:timeseries}
\end{figure}

The fitting procedure has been performed  for all the epochs corresponding to the measurements of PAMELA and AMS-02 resulting in time-series of best-fit values for the modulation parameter $K_0$ depicted in Fig.\ref{fig:timeseries}.
The vertical dashed line, along with the shaded area around it, represents the reversal phase, whose start and end periods were taken from the derivations in \citet{Sun_reversal_2013}.
The bottom panel illustrates the monthly and the smoothed SSN values, obtained by averaging over a 13-month period, computed by the SILSO/SIDC center at the \textit{Royal Observatory of Belgium} (Brussels, Belgium).
Notably, the propagation parameter exhibits a significant temporal dependency, which appears to correlate with solar activity. 
Specifically, it appears to peak during the unusually prolonged solar minimum of 2009–2010 \citep{Potgieter_2009_min} in A $< 0$ phase. Conversely, its minimum is observed during the solar maximum around 2013.

The $K_0$ parameter exhibits an anticorrelation with the smoothed SSN, which can be physically interpreted as faster cosmic ray diffusion within the heliosphere during quiet periods (high values of $K_0$), resulting in a milder attenuation of the LIS, thus leading to a higher flux of GCR in the GeV energy range. Decreased $K_0$ values signify slower CR diffusion, commonly observed during periods of intense solar activity when the modulation effect is pronounced. This pattern can also be understood qualitatively within the framework of the Force-Field approximation.
We also observe the effect of the Forbush decrease attributed to the anomalous solar activity of mid-2017 \citep{Dorman_2019}. It is worth noting that explosive solar events, such as flares and Coronal Mass Ejections (CMEs), induced spatially and temporally localized perturbations in the inner heliosphere, leading to short-term variations in the GCR flux (see Sect. \ref{sec:EMD}).

\subsection{Stochastic variations and smoothing}
\label{sec:EMD}
\begin{figure*}[t]
    \centering
    \includegraphics[width=0.85
\textwidth]{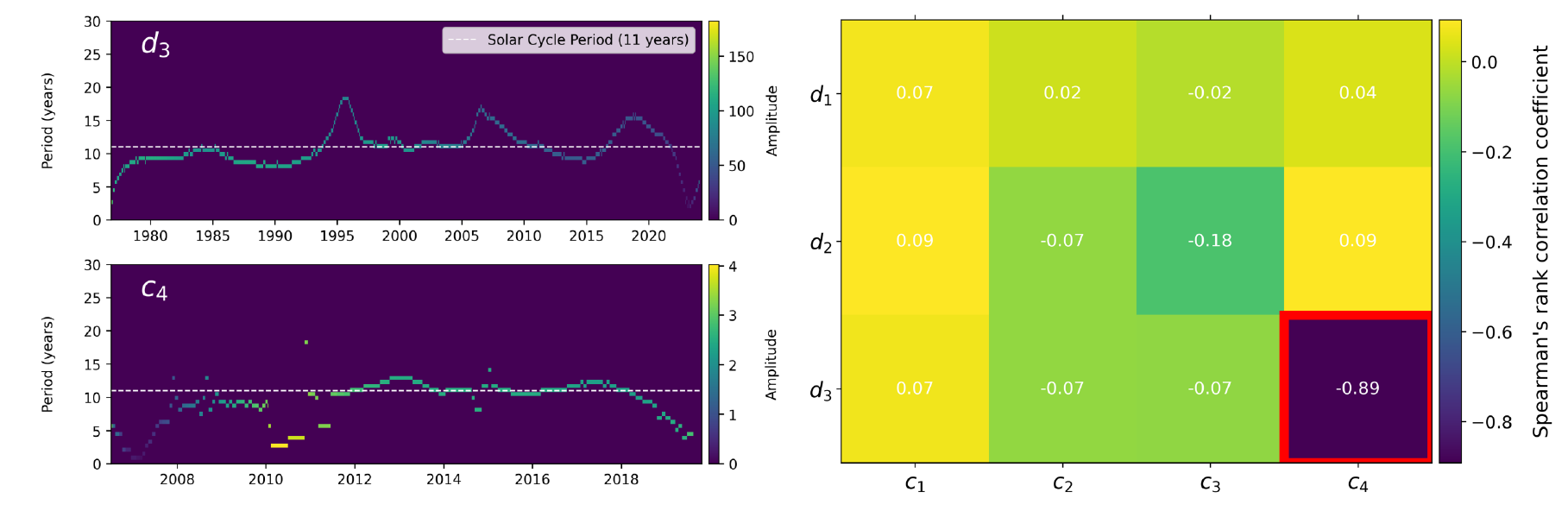}
\caption{Left: Hilbert spectrum of the $d_3$ mode of the solar proxy, \REV{and of $c_4$ from the modulation parameter, highlighting their shared $\sim$11-year periodicity over time}. Right: Mode-by-mode cross-correlation matrix between the modulation parameter $K_0(t)$ and the smoothed sunspot number, computed using Spearman's rank correlation coefficient. The red border marks the combination where the correlation is statistically significant, indicating that only $c_4$ shares the same long-term evolution as $d_3$.}
\label{fig:hht}
\end{figure*}

GCR fluxes are subject to a series of recurrent and non-recurrent variations. Precision measurements of daily proton fluxes performed by AMS-02 have shown that GCR fluxes exhibit recurrent variations with a period of 27 days, as well as shorter periods of 9 and 13.5 days, observed in 2016, each with peculiar rigidity dependencies \citep{AMS_PRL2021}. The 27-day periodicity is due to the rotational period of the Sun, while the other variations, as shown in \citet{CIR_Luo}, could be interpreted as the effect of prominent structures of compressed plasma in the solar wind, known as Corotating Interaction Regions. These regions form as a result of the compression of solar wind at the interface between fast and slow streams, or due to the latitudinal gradient caused by crossing the heliospheric current sheet.
Non-recurrent variations, in contrast, are associated with intense stochastic solar phenomena, such as solar flares or interplanetary CMEs, which significantly affect the model's parameters. As example CMEs can trigger a Forbush decrease, causing a sudden drop in GCR intensity near Earth \citep{Wang_2023,Romaneehsen}. These events can last for several days and can reduce GCR intensity under normal solar conditions by approximately 30\% to 40\%.
Given that the objective is to forecast the long-term GCR flux, we need to mitigate excessive sensitivity to stochastic short-term fluctuations that cannot be predicted simply through a solar proxy, as this might result in overfitting. Hence, before extracting cross-correlation functions, a filter was employed to generate a smoothed time-series of the modulation parameter, reducing the influence of short-term variations.
To tackle this task, we employed empirical mode decomposition (EMD) \citep{EMD}, leveraging its capability to extract meaningful components from raw data without necessitating predefined basis functions. EMD is an adaptive, data-driven algorithm that breaks down a signal into \textit{intrinsic mode functions} ($c$), providing a powerful tool for analyzing non-stationary and non-linear data. The fundamental concept behind EMD involves iteratively decomposing a signal into intrinsic mode functions, each representing a distinct oscillatory mode within the data.
The sifting algorithm scans the signal using a time-domain procedure designed to separate rapid dynamics in a time-series while progressively filtering out slower dynamics. This is achieved by iteratively subtracting the average of the signal’s upper and lower amplitude envelopes until the difference falls below a defined threshold, referred to as the sifting parameter. This stops the process and leads to the termination of the algorithm, isolating a $c$. The first intrinsic mode function is then subtracted from the original signal, and the process is repeated until no further modes can be identified, ultimately isolating the residual trend.
This decomposition methodology relies on the local characteristics of the signal, rendering EMD particularly effective for analyzing complex signals exhibiting varying frequencies and amplitudes over time.
Following the strategy described above, the modulation parameter is decomposed as follows:
\begin{equation}
    K_0(t) = \sum_{i=1}^{4} c_{i}(t) + r, 
    \label{eq:EMD}
\end{equation}
where $i = 1$ corresponds to the highest frequency mode, capturing the fastest oscillations in the signal, and $i = 4$ represents the lowest frequency mode for the selected sifting parameter. The term $r$ captures the long-term trend in the signal.
Given that our primary interest lies in capturing the long-term trend in the solar proxy, we applied the same decomposition strategy to the smoothed sunspot number timeseries ($S$), previously shifted by the time-dependent delay to take into account the plasma dynamic (see \ref{sec:cross_correlation}). Among the extracted modes, we identified the component denoted as $d_3$, which exhibits variability on a timescale of approximately 11 years, consistent with the solar cycle. To validate the frequency content of this mode, we examined its Hilbert spectrum (see Figure \ref{fig:hht}), which confirmed the expected frequency characteristics.
To isolate the most relevant IMFs from $K_0(t)$ for reconstructing its long-term evolution, we employed a correlation-based approach, comparing each mode with the long-term component of the solar proxy.
Specifically, we computed the Spearman correlation coefficient between each IMF and the selected solar cycle mode $d_3$. The Spearman coefficient was chosen for its robustness in capturing non-linear monotonic relationships. Our analysis revealed that only the mode $c_4$ exhibited a statistically significant correlation with the long-term component of the solar proxy. \REV{This finding is further supported by the Hilbert spectrum, shown in Figure \ref{fig:hht}, which highlights the shared frequency content between $c_4$ and $d_3$}. All remaining IMFs were interpreted as representing short-term fluctuations not associated with significant variability in the solar proxy and were excluded from the reconstruction.
Consequently, the smoothed time-series for the model parameter is given by $K_{s}(t) = c_4(t) + r$. Figure \ref{fig:timeseries} compares the original $K_0$ time-series with the smoothed version $K_{s}$.

\subsection{Cross-Correlations}
\label{sec:cross_correlation}
Previous studies, such as \citet{Wang_Chi2} and \citet{Fiandrini}, have attempted to establish direct correlations between solar proxies and GCR modulation parameters using proxies like the HMF magnitude $B_0$ and the smoothed SSN. These studies have shown significant discrepancies between the parameters obtained from flux data of PAMELA and AMS-02 across different solar periods (e.g., ascending/descending or defined HMF polarity phases).
This highlights the need to isolate correlation functions by splitting the phases of constant polarity, while excluding data from the magnetic reversal phase. The criteria outlined in \citet{Sun_reversal_2013} were used to identify and isolate this period.
Moreover, despite clear dependencies on the solar proxy being observed for the model parameter, modulation loops (hysteresis structures) are evident, as observed in \citet{Fiandrini}. To address the resulting double-valued behavior in the correlation function, applying a time lag is essential to account for the delayed response of GCR flux to solar variations.
This lag, as noted in \citet{Lag_General}, typically spans several months between solar proxies (such as monthly SSN, tilt angle, or HMF local magnitude) and the corresponding variations in neutron monitor rates or space-borne GCR measurements. 
Lag values can vary from 0 to 18 months depending on the epoch, solar cycle, datasets, or proxies used. Previous studies have also noted an odd-even cycle dependence, likely influenced by drift effects in heliospheric modulation \citep{Lag_BON}.
In this work, we correlate the delayed smoothed sunspot number $S(t-\tau)$ with the smoothed modulation parameters $K_{s}(t)$ where $\tau$ is the time lag derived from the following effective formula
\begin{equation}
\label{eq:lag}
\tau(t) = \tau_{M} + \tau_{A} \cdot \cos\left[\frac{2\pi}{T_{0}}(t-t_{p})\right]
\end{equation}

\begin{table}[h!]
\centering
\begin{tabular}{l c}
\hline
\textbf{Parameter} & \textbf{Value} \\
\hline
$\tau_M$ (months) & $9.82 \pm 0.42$ \\
$\tau_A$ (months) & $4.87 \pm 0.55$ \\
$T_0$ (years)     & $21.44 \pm 0.73$ \\
$t_P$ (years)     & $2.25 \pm 0.51$ \\
\hline
\end{tabular}
\vspace{0.5em}
\caption{\REV{{Values of the time-lag parameters used in Eq. \ref{eq:lag}, based on the formulation proposed in \citet{tomassetti}.}
}}
\label{tab2}
\end{table}

\begin{figure}[!hb]
\vspace{1em}
    \centering
    \includegraphics[width=0.45\textwidth,trim={0 50px 0 0}]{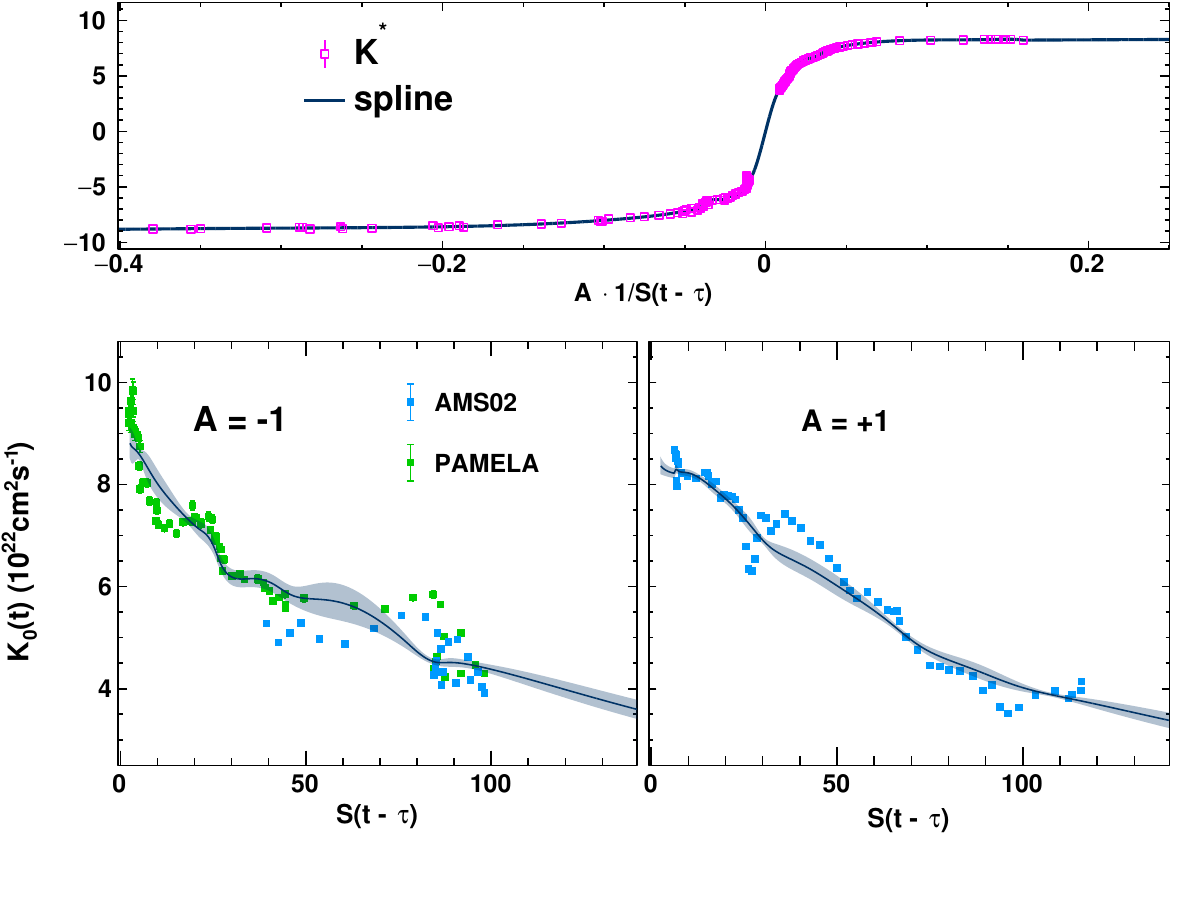}
\caption{The top row displays the spline function $\vec{f}$, shown as blue lines, for the model's modulation parameter obtained using the penalized spline fit defined in Eq. \ref{eq:loss_score}. The magenta data points correspond to the $K^{*}$ parameter derived from Eq. \ref{eqnarray:bias_trasform} plotted against $A/S(t\mathord{-}\tau)$. 
The bottom row displays the smooth cross-correlation function $\vec{f}$ during the negative and positive polarity phases, alongside the original $K_0(t)$ values for reference, plotted against the $S(t\mathord{-}\tau)$ proxy. These plots demonstrate how the long-term evolution of the model's modulation parameter is effectively captured by the derived cross-correlation functions.}
\label{fig:splines_global}
\end{figure}

This formula, proposed in \citet{tomassetti}, defines $\tau_A$ as the maximum amplitude of the variation, $T_0$ represents the oscillation period, $t_P$ represents the phase, and $\tau_{M}$ represents the average value around which the function oscillates. The parameters used in this work are related to the calibration dataset from IMP-8 and ACE, \REV{and are listed in Table~\ref{tab2}}.
Since no discontinuities are observed in the GCR fluxes during the reversal phase, the correlation function must transition smoothly between polarity phases.
Nevertheless, the limited coverage of solar activity provided by PAMELA and AMS-02 data constrains our ability to derive correlation functions generalizable to other solar cycles with higher $S$ values, particularly given the anomalous low intensity of the 24th solar cycle, where most of the data lays.
For these reasons, and given the effective nature of the model, as well as the lack of information regarding the functional form of the relationship between $K_0$ and $S$ (aside from the fact that $K_0$ is known to be positive and to decrease with increasing turbulence of the HMF), we set our goal to be as flexible as possible.
Instead of testing polynomial analytical forms, we performed a cubic B-spline fit to smoothly connect regions of opposite polarity $K_{s,-}$ and $K_{s,+}$ by defining a new proxy $A/S(t-\tau)$, where $A = \pm 1$ is the HMF polarity. This proxy was chosen because it takes values close to zero during periods of high solar activity to ensure a smoother transition in the high-$S$ extrapolation region, where no data are available.
The parameters are now defined as:
\begin{eqnarray}
\label{eqnarray:bias_trasform}
K^{*}_{+}(t) & = &K_{s,+}(t) \cdot A \\
K^{*}_{-}(t) & = & b + A \cdot K_{s,-}(t)    \nonumber
\end{eqnarray}
where $b$ represents an appropriate factor designed to ensure a smooth connection between the positive and negative branches, as depicted in Fig. \ref{fig:splines_global} (top row).
We tested various configurations for knot placement, including uniform and Chebyshev spacing, as well as a quantile-based strategy, which involves placing knots at the quantiles of the data to ensure they are distributed in a way that reflects the data’s distribution.
To achieve smooth functions, a penalized spline fit was performed. This involved fine-tuning the number of knots $N_{\text{knots}}$ and $b$, based on the configuration that minimizes the score s, which is defined by the following loss function:
\begin{equation} 
\label{eq:loss_score}
s = \sum_{i=1}^{n} \left( y_i - f(x_i) \right)^2 + \lambda \int f''(x)^2 dx
\end{equation}
where $x_i$ is the value of the new proxy and $y_i$ and $f(x)$ are respectively the value of the $K^{*}$ and the correlation function evaluated at $x_i$. Polynomial integration is particularly sensitive to edge effects (such as Runge's phenomenon) which can result in undesired oscillations on the spline function. To mitigate these issues and achieve a smoother spline, we apply a curvature penalty. For a detailed discussion of this method, refer to \citet{PENALTY}.
The first term of the loss function controls the agreement between the data and the spline fit, while the second term controls the smoothness of the spline. The best knot configuration was obtained using the quantile strategy described above.
The results of the spline procedure applied to $K^{*}$ are presented in Fig.\ref{fig:splines_global}, along with the derived correlation functions during defined polarity phases.
The uncertainty bands were derived using a fully statistical approach, considering that neither the EMD procedure, through which $K_{s}$ is obtained, nor the spline-fitting process naturally account for uncertainties.
Starting from the uncertainties of the modulation parameter shown in Fig.\ref{fig:timeseries}, which originate from the fitting procedure applied to the experimental data, we applied a bootstrap technique to propagate these uncertainties by generating Gaussian-simulated datasets, assuming each dataset to be independent.
For each simulated dataset, we repeated the smoothing and spline-fitting procedures described earlier, resulting in a set of correlation functions $\vec{f}_{sim}$. This allowed us to quantify how uncertainties in the data propagate through the entire analysis process.
For a given $S$, across both polarity phases, the resulting $\vec{f}_{sim}(S)$ followed a Gaussian-like distribution around the central value $\vec{f}(S)$. Therefore, the uncertainty for each $S$ was estimated as the standard deviation of this distribution, providing a measure of the spread around the central spline fit.

\subsection{Undefined polarity phase}
\label{sec:methods}
During the magnetic reversal phase, the polarity cannot be defined due to the complex dynamics of the HMF. Following \citet{Fiandrini} and \citet{Reversal_Jiang_2023}, we make use of a transition function $W$ to weight the contributions of positive and negative polarity for any epoch.
Thus, the reconstruction of the flux $J$ at kinetic energy $E_{k}$ and a specific epoch $t$ is achieved as a superposition of fluxes in positive and negative polarity states:
\begin{equation}
    J(E_{k},t) = J^-(E_{k}, {K_{0}}^{-}) W_t + J^+(E_{k}, {K_{0}}^{+})[1-W_t]
\end{equation}
where ${K_{0}}^{-}$ and ${K_{0}}^{+}$ represent the modulation parameter relative to specific polarity. The transition function is defined as

\begin{equation}
    W_t = \left[1+\exp\left(\text{sign}(c)\, \frac{t-T_{rev}}{\delta t}\right)\right]^{-1}
\end{equation}
where $\delta t \approx 3$ months, $c = \pm 1$ respectively for reversals during even and odd solar cycles, and $T_\text{rev}$ marks the epoch of polarity reversals.

\section{Results}
\label{sec:Results}
\begin{figure*}[t]
    \centering
    \includegraphics[width=0.9\textwidth,trim={0 15px 0 0}]{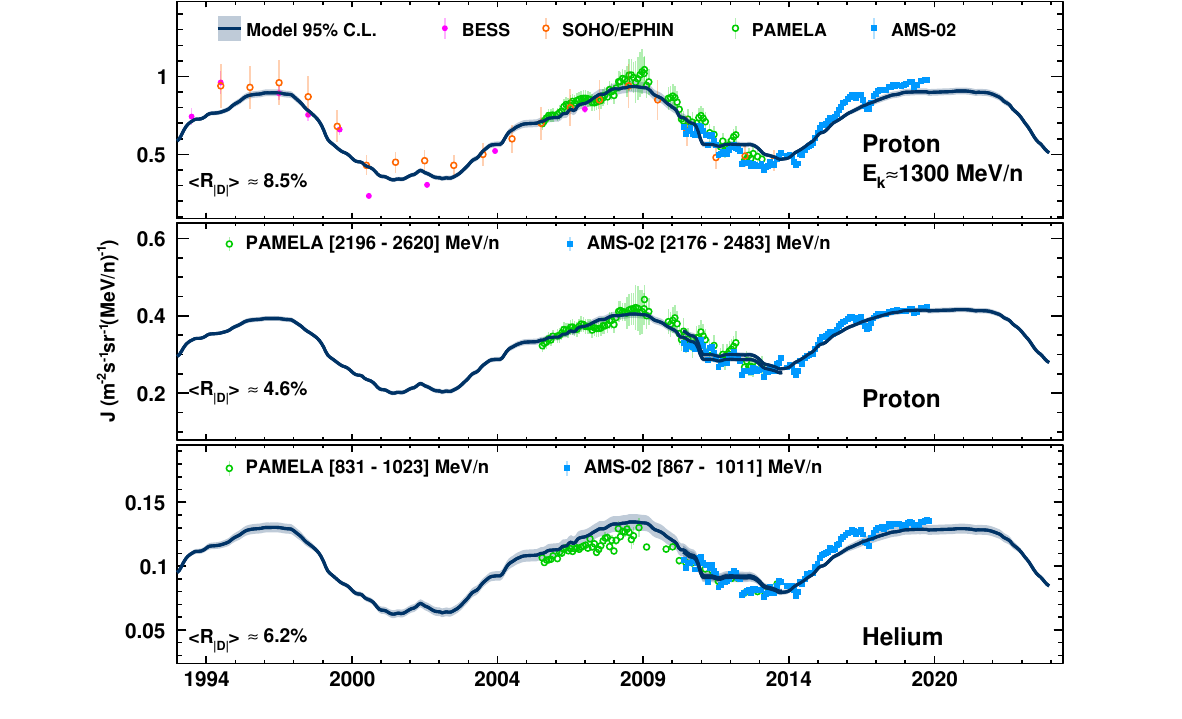}
\caption{Top plot: Results of the model evaluated at approximately $1300\,$MeV/n for protons using data from SOHO/EPHIN, BESS, PAMELA \citep{PAMELA_2013_protons, PAMELA_2018_protons}, and AMS-02 \citep{AMS_PRL2021} at the nearest energy bin to showcase reconstruction capabilities and accuracy in the past solar cycle.
Middle plot: Results of the model integrated into two reference energy bins close to $2\,$GeV/n, compared with Carrington rotation-averaged proton fluxes from PAMELA and Bartel’s rotation-averaged proton fluxes from AMS-02.
Bottom plot: Results of the integrated model in two energy bins, compared with Carrington rotation-averaged helium fluxes from PAMELA (\cite{PAMELA_2020_Helium}, \cite{PAMELA_2022_Helium}) and Bartel’s rotation-averaged helium fluxes from AMS-02 \citep{AMSPRL2022}. Each model is shown along with its uncertainty band. \REV{Each plot also displays the average of the relative error, calculated according to Equation \ref{eq:error}.}}
    \label{fig:long_term}
\end{figure*}

Once the smooth cross-correlation functions are determined, knowing the smoothed SSN value at epoch $t$ allows us to establish the parameter $K_0$ at epoch $t+\tau(t)$, enabling the forecast of CR fluxes.
It's noteworthy that this model was obtained using only the proton fluxes from PAMELA and AMS-02. 
However, as argued in \citet{NTPRL}, differences in the time variations of GCR fluxes can be attributed to variations in their spectral shapes (LIS) and the differing velocity dependence of the propagation parameters, which are influenced by the $A/Z$ ratio (mass-to-charge ratio) of each species.
Specifically, this dependence is embedded in the factor $\beta$, given by $\beta(P) = P/ \sqrt{P^2 + (m_p A/Z)^2}$, where $m_p$ is the rest mass of the proton, included in the definition of the diffusion coefficient Eq. \ref{eq:diffusion_coeff}.
Based on this argument, we assume that the parameter $K_0$ is universal, and the correlation functions can be applied to other CR species enabling forecasts if the LIS spectrum and isotopes abundances are known.
In Fig.\ref{fig:long_term}, the datasets for proton and helium fluxes at 1 AU are displayed at different reference energies alongside the forecasting model results and their uncertainty bands. The data used to compare come from AMS-02 \citep{AMS_PRL2021, AMSPRL2022}, PAMELA \citep{PAMELA_2020_Helium, PAMELA_2022_Helium}, while to extend the temporal comparison for protons we used data form other epochs SOHO \citep{SOHO} and BESS balloon experiment flights \citep{BESS97-2007, BESS93-97}.
The overall dataset spans approximately three solar cycles, covering two magnetic reversal phases that took place between September 2000 and May 2001, and between November 2012 and March 2014. 
\REV{To quantify the agreement between the model and the data, we compute the relative error, defined as:
\begin{equation}
R_{|D|} = \frac{| f_{\text{model}} - f_{\text{meas}} |}{f_{\text{meas}}}
\label{eq:error}
\end{equation}
where $f_\text{model}$ and $f_\text{meas}$ are the modeled and measured fluxes at a given epoch, respectively}. Despite measurements being taken at different energy and time intervals, ranging from monthly to multi-day intervals for BESS flights, a general agreement between the data and the model's results can be observed. 
This holds true even when considering the different levels of accuracy in the data from various experiments, particularly with the AMS-02 data, which has an experimental error of a few percent.
This agreement is particularly interesting for helium datasets, which serve as an independent validation test set since their fluxes were not used for calibration. This demonstrates that the universality of the modulation parameters, at least within the energy range considered, appears to be robust.
For a similar long-term evaluation of fluxes for heavier GCR species, such as Carbon and Oxygen, for which the LIS is known, we plan to utilize the latest AMS-02 data on the time variation of light-nuclei cosmic rays \citep{AMS_nuclei}, as it is currently the only operational instrument capable of providing such high-precision measurements.

\begin{figure}[!ht]
    \centering
    \includegraphics[width=0.5\textwidth]{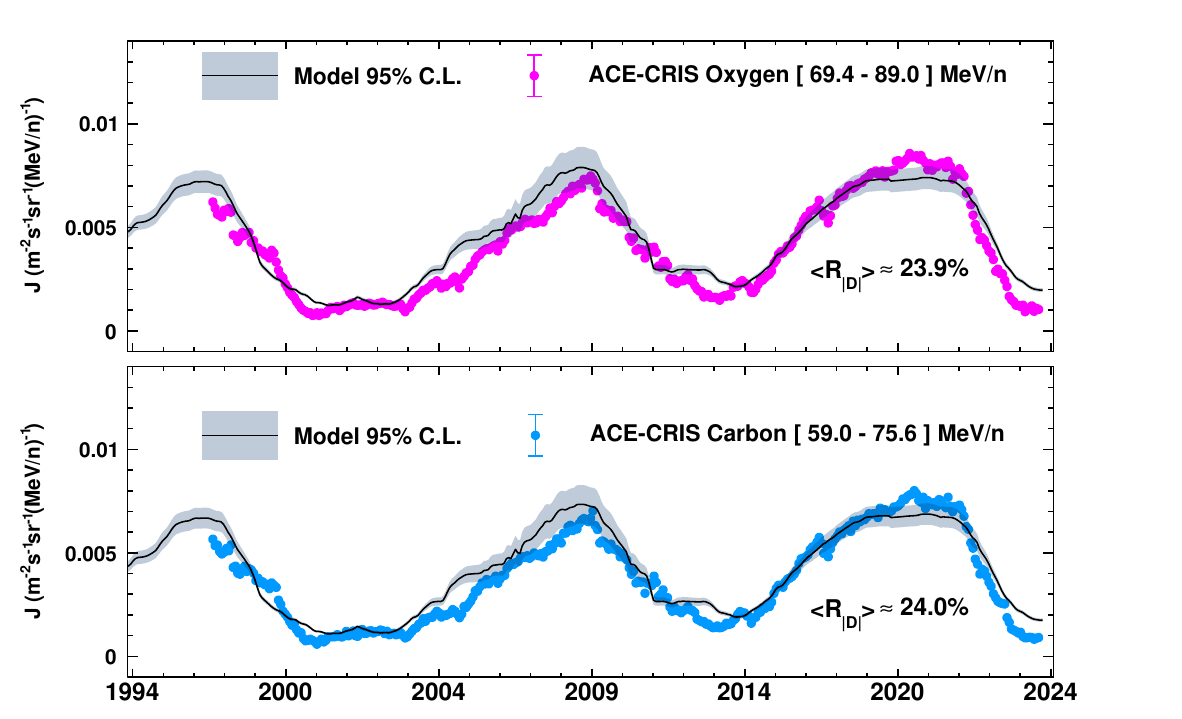}
    \caption{Results of the model evaluated at $80.4\,$MeV/n for Oxygen (magenta data-points) and $68.3\,$MeV/n for Carbon (azure data-points) shown as a solid blue line. The shaded band accounts for the model uncertainty; refer to the text for more details. The data are monthly averaged flux measurements from ACE/CRIS \citep{ACE/CRIS_first}. \REV{Each plot also displays the average of the relative error, calculated according to Equation \ref{eq:error}.}}
    \label{fig:ACE}
\end{figure}

Alternatively, a lower-energy evaluation of the model's performance can be conducted using the monthly fluxes from ACE for Carbon and Oxygen. Figure \ref{fig:ACE} shows the temporal variation of C and O fluxes along with the model reconstructions over three solar cycles. Despite the Carbon flux being extrapolated down to $68.3$ MeV/n and the Oxygen flux down to $80.4$ MeV/n, the overall reconstruction appears satisfactory, considering the significant approximations made during the development of the forecasting model, in this respect we plan to update it through a combined analysis of ACE and AMS-02 data, which will be the focus of future work.

Since the EMD smoothing procedure does not inherently account for uncertainties, we sought to quantify the uncertainty in our forecasting model using a statistically robust approach.  
To achieve this, we implemented a bootstrap technique: at each epoch where the smoothed sunspot number $S$, magnetic polarity, and kinetic energy $E_k$ are known, we generated a set of modulation parameter values by sampling from a Gaussian distribution. This distribution was defined using the previously derived correlation functions along with their associated uncertainty bands. For each simulated parameter, we computed the corresponding GCR flux at a reference energy using the numerical model, and we interpreted the standard deviation of the resulting flux distribution as the uncertainty. This procedure was repeated for both magnetic polarities and for varying values of $S$ and $E_k$.


\section{Discussion and Conclusions}
\label{sec:conclusion}
Forecasting cosmic ray fluxes across various nuclear species is crucial for properly planning future space missions and accurately assessing astronaut health risks associated with such endeavors. The recent GCR flux measurements provided by PAMELA and AMS-02, which will continue to supply data until the ISS is decommissioned, are crucial for investigating propagation phenomena.
In this paper, we present calculations for the interstellar fluxes of GCR nuclei using a revised Force-Field approach, integrating low-energy data from the Voyager missions with high-energy data from AMS-02.
Several simplifications were made in solving the Parker equation using a radial approximation, neglecting the tensor nature of cosmic ray diffusion, anisotropic diffusion effects, and curvature drift. The heliosphere was simplified by ignoring latitudinal solar wind variations, termination shock effects, and the heliospheric current sheet. Including these factors would increase model complexity and computational cost. In this respect, the model's parameter should be viewed as effective, averaged value rather than direct reflection of heliospheric conditions, and its temporal evolution should be interpreted accordingly.
In our model, time dependence is handled by providing a time-series of steady-state solutions for proton fluxes, based on time-varying input parameter. This approach is valid as long as cosmic ray transport timescales remain shorter than changes in solar activity. To ensure accuracy, the model parameters were calibrated using AMS-02 and PAMELA monthly proton data, covering multiple solar phases and a wide energy range. For shorter timescales (e.g., daily) or lower energies, a potential extension of this work is to explore a time-dependent solution to the Parker equation.
The reconstructed GCR flux, derived from our fitting process, demonstrates robust agreement with observed data across various solar activity phases, affirming the reliability of our model.
Our analysis highlights that the time-series of model parameter offer valuable insights into its evolution with solar activity proxies, such as the smoothed sunspot number. Specifically, we observe a notable long-term dependence in the modulation parameter indicating an evolving turbulence regime that tracks the solar cycle.
Moreover, our investigation into the correlation between solar and modulation parameter reveals charge-sign-dependent features in solar modulation, including distinct patterns during different polarity phases.
To extract this correlation, given our aim to capture the long-term evolution of GCR flux, we utilized a penalized spline fit preceded by a filtering process based on Empirical Mode Decomposition. This filtering step is crucial for removing short-term components associated with solar activity stochastic fluctuations, linked to eruptive events and highly perturbed conditions, which cannot be effectively captured solely by the smoothed SSN.
In this scenario, we considered the time lag, modeled with an empirical formula, between solar activity and the GCR flux observed at Earth, which allow us to make flux forecasts.
As demonstrated in the last section, our model exhibits impressive accuracy in flux reconstructions and predictive capabilities, despite the approximations assumed in the calculations and the limited dataset used for the calibration. This result has been validated by comparisons with data from various experiments across different energy intervals, epochs, and GCR species.
As a potential future application, given that we have developed correlation functions for specific polarities, we could leverage the LIS for antiprotons and, more broadly, for antinuclei to predict their fluxes. Furthermore, with the availability of high-energy resolved GCR fluxes, it will be feasible to explore the possibility of calibrating based on specific phases of the solar cycle, particularly the ascending and descending phases, to enhance the model's accuracy. 
We are already working on an extension of the current solar modulation model, providing a better parameterization of the flux by incorporating the latest AMS-02 nuclei data \citep{AMS_nuclei} and ACE low-energy data. These improvements will be detailed in future publications.
Therefore, our model shows significant promise for applications related to space weather forecasting and predicting absorbed doses for astronauts. We are currently developing a model to assess the radiation doses absorbed by astronauts both outside and inside the geomagnetic shield. To do so, we include detailed calculations of the rigidity cutoff, which will be presented in a dedicated publication.
Additionally, in the long-term scale, we could take advantage of SSN forecasts for future solar cycles as input to our model, thereby extending its predictive capabilities \citep{SSN_future_Penza_2021, Asikainen2023}.
For further enhancements, we aim to refine the consideration of lag dependence on solar activity parameters, such as solar wind speed or HMF polarity. Additionally, improvements in LIS refinement and exploration of radial dependence in the parameter $K$, as well as higher-dimensional spatial numerical calculations, are areas of interest that we are considering for future development.


\section*{Acknowledgments}
We acknowledge support from ASI under ASI-INAF 2020-35 HH.0 (CAESAR), ASI-UniPG 2019-2-HH.0, ASI-INFN 2019-19-HH.0, its amendments. No. 2021-43-HH.0, and the Italian Ministry of University and Research (MUR) through the program “Dipartimenti di Eccellenza 2023-2027”.
We also acknowledge support from FCT under grant 2024.00992.CERN, Portugal.



\end{document}